\documentclass{iopart}
\usepackage{graphicx}
\usepackage{cite}
\usepackage{iopams}

\begin{document}

\title[Nonlinear Schr\"odinger equation for a
$\cal PT$ symmetric delta-functions double well]
{Nonlinear Schr\"odinger equation for a
  $\cal PT$ symmetric delta-functions double well
}

\author{Holger Cartarius,  Daniel Haag, Dennis Dast, G\"unter Wunner}
\address{Institut f\"ur Theoretische Physik, Universit\"at Stuttgart,
         Pfaffenwaldring 57, 70\,569 Stuttgart, Germany}
\ead{wunner@itp1.uni-stuttgart.de}

\begin{abstract}
The time-independent nonlinear Schr\"odinger equation is solved 
for two attractive delta-function shaped potential wells 
where an imaginary loss 
term is added in one well, and a gain term of the same size but with
opposite sign in the other. We show that for vanishing nonlinearity
the model captures all the features known from studies of $\cal PT$
symmetric optical wave guides, e.g., the coalescence of modes 
in an exceptional point at a critical value of the loss/gain parameter, and
the breaking of $\cal PT$ symmetry beyond. With the nonlinearity present, the 
equation is a model for a Bose-Einstein condensate with loss and gain 
in a double well potential. We find that the nonlinear Hamiltonian picks
as stationary eigenstates exactly such solutions which render the nonlinear
Hamiltonian itself $\cal PT$ symmetric, but observe coalescence 
and bifurcation scenarios different from those known from 
linear $\cal PT$ symmetric Hamiltonians.
\end{abstract}

\submitto{\JPA}
\maketitle

\section{\label{sec:intro} Introduction}

The observation of parity-time ($\cal PT$) symmetry in experiments
with structured optical waveguides \cite{guo09,rueter10} marked a remarkable 
triumph for the mathematical
theory of non-Hermitian Hamiltonians having $\cal PT$ symmetry, initiated
by Bender and Boettcher \cite{bender98} in 1999. The experiments
exploit the optical-quantum mechanical  
analogy that the wave equation for the transverse electric
field mode is formally equivalent to the one-dimensional Schr\"odinger equation
when the potential is identified with $V(x) = -k^2n^2(x)$  and the energy
eigenvalue with $E = -\beta^2$, where $k = \omega/c$ is the vacuum wave number 
and $\beta$ the propagation constant of the mode. A $\cal PT$ symmetric
experiment is realized by introducing loss and gain terms 
in the complex index of refraction $n(x)$. 

A quantum analogue to
the optical experiments would be a Bose-Einstein condensate in a double-well
potential where loss and gain is realized by removing atoms from one well
and pumping in atoms in the other, as has been suggested by
Klaiman et al. \cite{klaiman08a}.
A complication arises from the interaction between the condensate atoms.
In the simplest case this is s-wave scattering, which yields an additional
(real-valued) potential term proportional to $|\psi|^2$, and turns the 
non-Hermitian Hamiltonian into a nonlinear one. For the Hamiltonian to
be $\cal PT$ symmetric this term has to be a symmetric function of the 
coordinates, which cannot be guaranteed from the outset. It will be the
purpose of this paper to analyze whether or not the nonlinearity 
destroys the $\cal PT$ symmetry of the quantum system. 

We will do this, in the spirit of a model calculation, 
by considering the  
situation where the double well is idealized by two attractive 
delta-function potential wells, with loss added in one trap and gain in the 
other. We will demonstrate that the stationary solutions of the
Gross-Pitaevskii equation indeed preserve the $\cal PT$ symmetry
of the nonlinear Hamiltonian, and merge in a branch point at some 
critical value of the 
loss and gain, beyond which the $\cal PT$ symmetry is broken. 

Another model which simulates the physical situation of a Bose-Einstein condensate in a  
double well with loss and gain
has been investigated by Graefe et al. 
\cite{graefe08a, graefe08b, graefe10} in the framework of 
a two-mode Bose-Hubbard-type $\cal PT$ symmetric Hamiltonian.
In the model the dynamics of the quantum system is mapped
on that of a (formal) spin vector whose motion on the surface of
the Bloch sphere can be analyzed in terms of classical nonlinear dynamics.
As an optical analogue, in the  two-mode approximation Ramezani et al. 
\cite{ramezani10} have recently looked at a mathematical model of a 
$\cal PT$ symmetric coupled dual waveguide arrangement with Kerr nonlinearity. 
It will be interesting to see which features of these models
are recovered when actually solving the nonlinear $\cal PT$ symmetric 
Gross-Pitaevskii equation.

Delta-function potentials are popular  as model 
systems in the literature  since they allow for analytic or partially analytic solutions 
but are flexible enough to provide insight into characteristic phenomena of more
complex physical situations. For example,
resonances and the decay behaviour of a Bose-Einstein condensate in a double delta-shell potential were analyzed by 
Rapedius and Korsch \cite{rapedius09} with the aim of modelling features
typical of double-well potentials \cite{theocharis06}. In the same vein,
Witthaut et al. \cite{witthaut08} have studied the nonlinear Schr\"odinger
equation for a delta-functions comb to gain insight into the 
properties of nonlinear stationary states of periodic potentials. 
For a system where a real delta-function potential is 
augmented by a $\cal PT$ symmetric pair of delta-functions
with imaginary coefficients, bound states 
and scattering wave functions have been calculated by Jones \cite{jones08}.
The paper was devoted to the quasi-Hermitian
analysis of the problem, and no nonlinearity was present. Jakubsk\'y and Znojil \cite{jakubsky05} have considered
the explicitly solvable model of a particle exposed to two imaginary $\cal PT$ 
delta-function potentials in an infinitely high square well, and determined 
the energy spectrum, but also without nonlinearity. 
A large body of literature of course exists 
on the properties of solitons and Bose-Einstein condensates in periodic 
optical and nonlinear lattices with $\cal PT$ symmetry 
and their nonlinear optical analogues (see, e.g.,
\cite{abdullaev07a,abdullaev07b,makris08,musslimani08a,musslimani08b,
abdullaev10,bludov10,makris11,abdullaev11}). 
But to the 
best of our knowledge the basic problem of the nonlinear 
Schr\"odinger equation with two $\cal PT$ symmetric
delta-function double wells has not been treated so far. 

\section{Delta-function traps model}

We consider a Bose-Einstein condensate trapped in two delta-function
shaped potential wells located at $x = \pm a/2$, where from one well
condensate atoms are removed and the same amount of atoms is added
in the other. At sufficiently low temperatures stationary
solutions for the condensate wave function are then described in a mean-field
approach by the nonlinear Schr\"odinger equation, or Gross-Pitaevskii
equation (see, e.g., \cite{Pitaevskii03a}),
\begin{eqnarray}\label{GPEdimless}
-\Psi^{\prime\prime}(x) &-&\left[(1+i\gamma)\delta(x+a/2) + (1-i\gamma)\delta(x-a/2)\right]\Psi(x) \nonumber \\
&-& g|\Psi(x)|^2 \Psi(x)
= -\kappa^2 \Psi(x)\, ,
\end{eqnarray}
where $g$ is the nonlinear interaction strength, and the real-valued parameter
$\gamma$ determines the size of the gain and loss terms.
Units have been chosen in such a way that the strength
of the real part  of the delta-function potentials  is normalized to unity.
(The relation to physical quantities is given in the Appendix.)
For stationary solutions the eigenvalues $\kappa$ will be real,
but since we are interested in the complete eigenvalue spectrum,
including complex eigenvalues, we will quite generally search for solutions 
with $\kappa \in {\mathbb {C}}, \; {\rm Re}(\kappa) > 0$. 
While the delta-function
potentials are $\cal PT$ symmetric, it is not clear a priori that 
the equation itself is $\cal PT$ symmetric since this requires
the nonlinear potential term to be a symmetric function of $x$.

To be in a position to study the effects that emerge when the nonlinearity 
is turned on we first consider the case where the nonlinearity
is absent, and then proceed to values $g >  0$.

\subsection{Delta-functions trap model for vanishing nonlinearity}

For $g = 0$ the bound-state wave function has the form:
\[
\Psi(x) = \left\{ \begin{array}{l@{\quad:\quad}l}
A\, e^{\kappa x} & x < -b \\
C\, e^{\kappa x} + D\, e^{-\kappa x} & -b < x < b \\ 
B\, e^{-\kappa x} & b < x
\end{array} \right.
\]
Applying the continuity conditions leads to the system of 
linear equations (with the
abbreviation $\kappa_0 = 1 + i \gamma$),
\begin{eqnarray}
\kappa_0 {\rm e}^{-\kappa b} \, C+ (\kappa_0-2\kappa){\rm e}^{\kappa b}D &=& 0 \;,
\label{cd1}
\\
 (\kappa_0^\ast-2\kappa){\rm e}^{\kappa b}\, C + \kappa_0^\ast {\rm e}^{-\kappa b}\,D
&=& 0 \;.\label{cd2}
\end{eqnarray}
The eigenvalues $\kappa$ are obtained numerically by finding the roots of 
the complex secular equation
\begin{equation}
(1+\gamma^2) {\rm e}^{-2\kappa a} - (1 + \gamma^2 + 4\kappa^2 -4 \kappa)=0 \,.
\end{equation}
It depends on two parameters, the distance $a$ of the traps and the
strength $\gamma$ of the loss/gain terms.

\begin{figure}[tb]
{\begin{center}
  \includegraphics[width=0.6\textwidth]{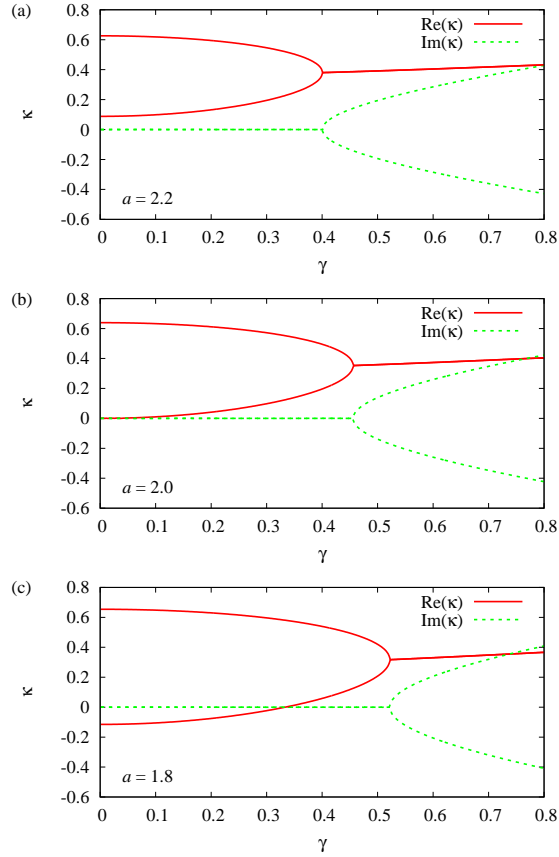}
\end{center}
\caption{\label{fig:1}
Eigenvalues $\kappa$ of equation (\ref{GPEdimless}) for vanishing
nonlinearity as functions of
the size of the loss/gain parameter $\gamma$ for different distances of 
the delta-function traps: (a) $a=2.2$, (b)  $a=2.0$. (c) $a=1.8$.
}}
\end{figure}

In figure~\ref{fig:1} the eigenvalues $\kappa$ are plotted as functions of the 
strength of the loss/gain parameter $\gamma$ for different distances of 
the traps. 
It can be seen that the solutions exhibit the behaviour typical
of $\cal PT$ symmetric non-Hermitian Hamiltonians: up to a critical
value of the loss and gain parameter $\gamma$ there exist two real
eigenvalues $\kappa$. 
Since we have included a minus sign on the right-hand side of the
eigenvalue equation (\ref{GPEdimless}), in the range of real
eigenvalues the upper branch belongs to the ground state and the 
lower branch to an excited state. 
At the critical value -- the branch point
(or exceptional point) -- the eigenvalues and the eigenfunctions merge.
Beyond the branch point the eigenvalues become complex, and complex conjugate.
The time-dependence of the eigenstates of (\ref{GPEdimless})
with $g= 0$ is given by $\exp(i\kappa^2\,t)
= \exp(i(k_{\rm r}^2-k_{\rm i}^2)t) \exp(-2k_{\rm r}k_{\rm i}t)$ (with $k_{\rm r}$ and
$k_{\rm i}$ the real and imaginary part of $\kappa$), hence the mode with
$k_{\rm i} >0$ decays, and the mode with $k_{\rm i} < 0$ grows.

Of course at $\gamma=0$ the two states
turn into the well-known $\cosh(\kappa x)$ (ground state) and $\sinh(\kappa x)$ 
(excited state) textbook solutions of  (\ref{GPEdimless})
between the traps. It is also
well-known that for $\gamma = 0$ the excited state disappears at a 
distance of $a = 2$. From figure~\ref{fig:1} (b) we see, however, that for $a = 2$
an excited state is born as soon as loss and gain is switched on.
The Figure also shows that for distances below $a=2$ (cf. figure~\ref{fig:1} (c))
the excited state
is born when loss and gain exceed a finite threshold value.
 
The behaviour of the eigenvalues shown in figure~\ref{fig:1}
is in complete analogy with that found by Klaiman et al.
in their study of two micro waveguides with loss and gain (cf. figure 2
in Ref. \cite{klaiman08a}). Their calculation was performed for
a fixed distance of the  waveguides. It would be interesting to
investigate whether or not the excited mode also vanishes
when the distance of the waveguides is decreased.

Below the branch point the wave functions possess $\cal PT$ symmetry,
$\Psi^\ast(x) = \Psi(-x)$. In our model this can easily be seen
since for real $\kappa$ the complex conjugate of  (\ref{cd1}) 
is identical to  (\ref{cd2}) so that the pairs $(C, D)$ 
and $(D^\ast, C^\ast)$ fulfill the same equation, and, hence $D=C^\ast$. 
From the continuity condition then follows that $A = C + C^\ast e^{2\kappa b}$
and $B= C e^{2\kappa b} + C^\ast = A^\ast$. Note that quite generally
the $\cal PT$ symmetry of a wave function implies that 
its modulus is a symmetric function, since $\Psi^\ast(x)\Psi(x) =
\Psi(-x)\Psi^\ast(-x)$.

\begin{figure}[tb]
\begin{center}
\includegraphics[width=0.6\textwidth]{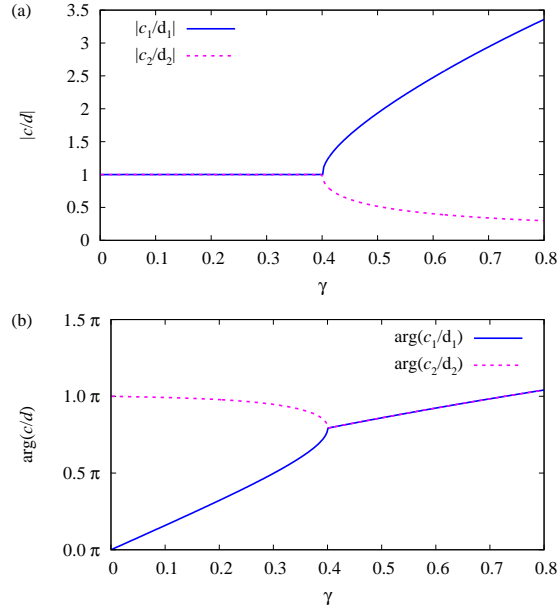}
\end{center}
\caption{\label{fig:2}
Real parts, (a), and imaginary parts, (b), of the ratio of the coefficients
of the wave functions in the region between the delta-function traps
for vanishing nonlinearity and a distance of $a = 2.2$ as functions of
the size of the loss/gain parameter $\gamma$. For values of $\gamma$ below
the bifurcation point
the solid lines correspond
to the ground state, 
the dashed lines to the excited state.
}
\end{figure}

Figure~\ref{fig:2} shows, for $a = 2.2$, the
ratio of the coefficients $C$ and $D$ of the wave functions 
in dependence of the loss/gain parameter $\gamma$. 
What is shown is the evolution of the modulus of the ratio and
its phase for the ground state and the excited state.
Since below the branch point we have $D=C^\ast= |C| \exp(-i\varphi)$, the modulus of the ratio is 1, and the phase is twice the phase $\varphi$ of the
coefficient $C$. As it must be, at $\gamma = 0$ the phase starts with a 
value of 0 for the (symmetric) ground state and $\pi$ for the 
(antisymmetric) excited state.
While the phase for the ground state is quickly rotated away 
from its initial value, the phase of the ratio belonging to the excited state
only slowly deviates from $\pi$ up to the branch point.
At the branch point the modulus and
the phase of $D/C$ coincide. Beyond the branch point they quickly evolve
away from the values they have in the $\cal PT$ symmetric regime.

\begin{figure}[tb]
\begin{center}
\includegraphics[width=0.6\textwidth]{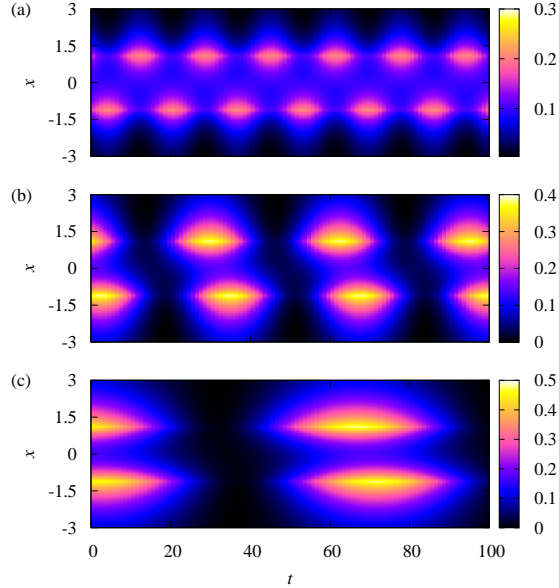}
\end{center}
\caption{\label{fig:time_evolution}
Time evolution of a superposition of the two eigenstates for different values
of the loss/time parameter: (a) $\gamma = 0$: (b) $\gamma = 0.35$, 
(c) $\gamma = 0.39$, close to the exceptional point. 
 }
\end{figure}

It is instructive to look at the time dependence of a superposition of the 
two eigenstates in the $\cal PT$ symmetric regime. For different
values of the loss/gain parameter 
figure~\ref{fig:time_evolution} shows the time evolution of 
\begin{equation}\label{timeevolutionlinear}
|\Psi(x,t)|^2 = |\psi_1(x)\exp(i \kappa_1^2 t) + \psi_2(x)\exp(i \kappa_2^2 t)|^2.
\end{equation}
Without the loss/gain term the probability density oscillates between the two delta-function wells. For finite gain and loss the beat frequency decreases
and the oscillation is deformed, and for  a loss/gain parameter close
to the branch point the probability density on both sides is almost identical
and pulsates.  
This behaviour completely agrees with the time evolution of the 
power distribution 
for a propagating electric sum field consisting of two guided modes calculated
by Klaiman et al. (cf. figure 3 in Ref. \cite{klaiman08a}). 

Thus we find that the simple quantum mechanical model
studied in this section already captures, for both 
eigenvalues and wave functions, all the effects
of a $\cal PT$ symmetric wave guide configuration in optics.

\subsection{Delta-functions trap model with nonlinearity
}
\subsubsection{Numerical method}
For $g\ne 0$ we have solved the Gross-Pitaevskii equation  
(\ref{GPEdimless}) numerically using a procedure in which 
the energy eigenvalues are found by a five-dimensional numerical root search. 
The free parameters which have to be adjusted in such a way that a
physically meaningful wave function is obtained are the eigenvalue $\kappa$ 
as well as initial conditions for the wave
function and its derivative. Since the overall phase is arbitrary we can
choose it such that $\Psi(0)$ is a real number. Therefore five 
real
parameters remain, viz. the real part of $\Psi(0)$, 
and the real and imaginary parts of both $\Psi^\prime(0)$
and $\kappa$. Physically relevant wave functions must be square integrable and
normalized. The normalization is important since the
Gross-Pitaevskii equation is nonlinear and the norm influences the form
of the
Hamiltonian. This gives in total five conditions which have to be fulfilled:
The real and imaginary parts of $\Psi$ must vanish for $x \to \pm \infty$, and
the norm of the wave function must fulfill $||\psi|| - 1 = 0$.

Outside the delta-function traps the Gross-Pitaevskii equation 
(\ref{GPEdimless}) coincides with the free nonlinear Schr\"odinger 
equation, which has well known real
solutions in terms of Jacobi elliptic functions 
(cf., e.g., \cite{witthaut08,carr00,carr01}). 
The function which solves the equation in the ranges $|x| > a/2$ for 
the attractive nonlinearity considered here and decays to zero 
for $|x| \to \infty$ is ${\rm cn}(\kappa x,1) =  
{\rm sech}(\kappa x)= 1/{\rm cosh}(\kappa x)$. We find that
once the correct eigenvalues and eigenfunctions are obtained 
our numerical wave functions exactly show this behaviour. 

\subsubsection{Eigenvalues}

\begin{figure}[tb]
\begin{center}
\includegraphics[width=0.6\textwidth]{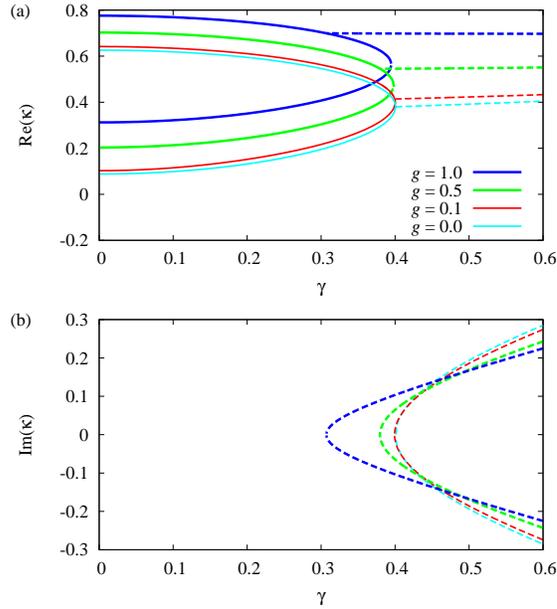}
\end{center}
\caption{\label{fig:3}
Real parts (a) and imaginary parts (unless zero) (b) of the 
eigenvalues $\kappa$ of the full nonlinear equation (\ref{GPEdimless})  
as functions of the loss/gain parameter $\gamma$ for $a=2.2$
for the four values of the nonlinearity parameter (in (a) from bottom to top)
$g = 0.0,~ 0.1,~ 0.5,$ and $~1.0$.
Solid lines denote purely real eigenvalues, dashed lines
complex conjugate eigenvalues.
For $g\ne 0$ the complex conjugate eigenvalues bifurcate
from the ground state branch before the branch point where the two real
solutions coalesce.}
\end{figure}

Figure~\ref{fig:3} shows the results for the real and imaginary parts of the
eigenvalues $\kappa$ calculated
for a trap distance of $a =2.2$ and different values of the nonlinearity 
parameter as functions of $\gamma$. The results for the case $g=0$ are
shown for comparison. It can be recognized that even with the nonlinearity
present there always exist two branches of real eigenvalues, up to 
critical values 
$\gamma_{\rm cr}$, where the two branches merge. It can also be seen
that branches of two complex conjugate eigenvalues also appear, 
but it comes as a surprise that these are born, not at $\gamma_{\rm cr}$, 
but at smaller values of $\gamma$ where they bifurcate from the real eigenvalue branch of 
the ground state. As the nonlinearity is increased, 
the bifurcation points are shifted to smaller values of $\gamma$, while
the branches of the real eigenvalues practically retain their form and 
are shifted upwards.
We therefore find that for nonvanishing nonlinearity 
there always exists a range of $\gamma$ values where
two real and two complex eigenvalues coexist. 

This behaviour obviously is a new effect, caused by the nonlinearity 
in the Gross-Pitaevskii equation. In fact the different branches can be 
continuously transformed into each other: when, for fixed $\gamma$, 
we increase the nonlinearity from $g=0$ in small
steps to $g=1$ and always take the solution of the previous step as input
for the root search in the next step, the branch of complex conjugate
eigenvalues which exists for $g=0$ is continuously transferred to the
branch of complex eigenvalues for $g=1$ shown in figure~\ref{fig:3}.
Likewise, starting from the branch of complex conjugate
eigenvalues for $g=1$ at any value of $\gamma$ in the range where 
four solutions exist and decreasing $g$ in small steps to $g=0$
we end up on the branch belonging to the real eigenvalue of the ground state.

At this point it is useful to compare  with the results
of the investigations of a $\cal PT$ symmetric two-mode Bose-Hubbard 
Hamiltonian with loss and gain by Graefe et al. \cite{graefe10}.
An eigenenergy spectrum with a structure similar to the one shown
in figure~\ref{fig:3} also 
appeared in their calculations (see figure~13 in \cite{graefe10}).
In the model, stationary states correspond to fixed points of the motion
of a vector on the Bloch sphere,
whose types can be classified according to the eigenvalues of the Jacobian 
matrix.
In the region where only two real eigenvalues exist the fixed points
can be identified as centres, while in the region with four 
eigenvalues the fixed points correspond to a centre and a saddle point,
and a sink and a source.
The centre and saddle point collide at the branch point and vanish.

This behaviour is in beautiful agreement with the results shown 
in figure~\ref{fig:3}. A stability analysis of the wave functions shows
that up to the bifurcation point both the ground state and the excited
state are stable, they correspond to the two centres. Beyond the bifurcation
point the excited state remains stable (a centre) while the ground
state becomes unstable (a saddle point). Out of the wave functions 
belonging to the two complex conjugate eigenvalues one decays and the
other grows, corresponding to the sink and the source, respectively.
It is gratifying that the two so different approaches qualitatively
yield the same behaviour.
It can be concluded that quite generally
the familiar branching scheme for linear $\cal PT$ symmetric 
Hamiltonians will be changed into a scheme of the type shown in
figure~\ref{fig:3} if a nonlinearity is added to the Hamiltonian.

\begin{figure}[tb]
\begin{center}
\includegraphics[width=1\textwidth]{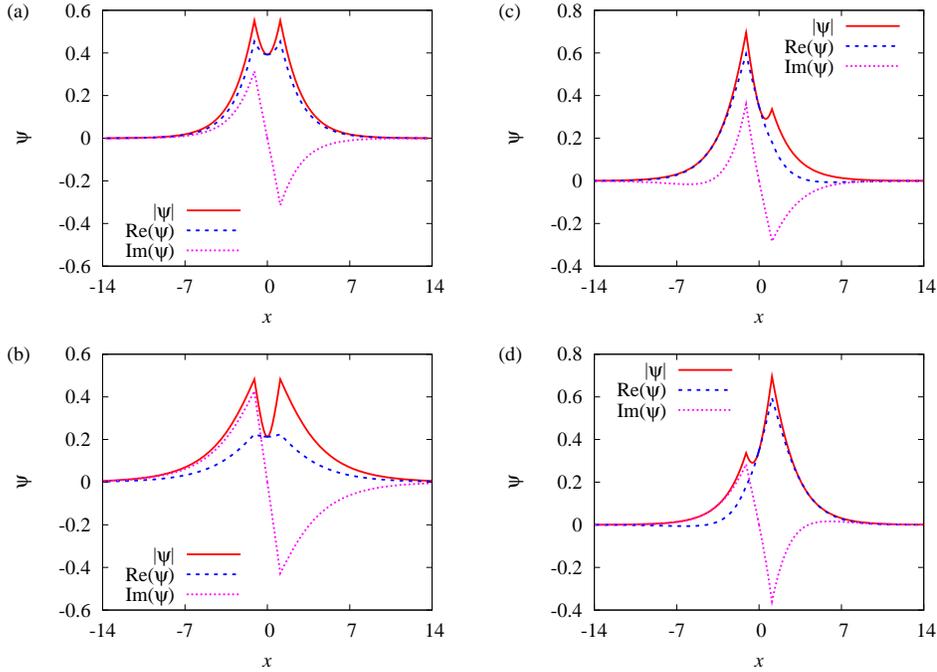}
\end{center}
\caption{\label{fig:4a}
Real and imaginary parts and moduli of wave functions of
the eigenstates of the nonlinear
Hamiltonian in  (\ref{GPEdimless}) for $g=0.5$, $a=2.2$.
(a) Ground state and (b) excited state for $\gamma =0.35$, (c) 
solution with imaginary part of $\kappa > 0$ and (d) imaginary part of $\kappa <0$  for $\gamma = 0.5$. In (a) and (b), the  wave functions are $\cal PT$ 
symmetric, and the moduli 
are symmetric functions, producing the $\cal PT$ symmetry of the total 
nonlinear Hamiltonian. The wave functions in (c) and (d) are no longer
$\cal PT$ symmetric, their moduli are not
symmetric functions, and the $\cal PT$ symmetry also 
of the nonlinear Hamiltonian is broken. 
 }
\end{figure}

\subsubsection{Eigenfunctions}

In figure~\ref{fig:4a} we present examples of wave functions of the eigenstates of the
nonlinear Schr\"odinger equation  (\ref{GPEdimless}) 
determined numerically for $g=0.5$ and $a=2.2$.
The results for the ground state and the excited state are shown 
in figure~\ref{fig:4a}~(a) and (b) for $\gamma = 0.35$, below the the critical 
value $\gamma_{\rm cr}\approx 0.4$.
The fact that in the numerical calculation the global phase of the complex wave function
was fixed in such a way that $\Psi(0)$ is real makes it 
particularly easy to recognize
the $\cal PT$ symmetry of the two solutions, since their 
real parts become even functions and their imaginary parts odd functions
of $x$. From the $\cal PT$ symmetry of the wave function follows that the 
modulus, also shown in figure~\ref{fig:4a}, is an even function, 
and with it
the nonlinear term in  (\ref{GPEdimless}). We therefore have the 
important result 
that the nonlinear Schr\"odinger equation with loss and gain 
selects as eigenfunctions exactly those states
in Hilbert space which render the nonlinear Hamiltonian $\cal PT$ symmetric!
In the ground state, which emerges from the symmetric real wave function
for $\gamma=g=0$, the symmetric contribution from the real part still
dominates, while for the excited state, which originates from
the antisymmetric solution for $\gamma=g=0$, the antisymmetric
contribution from the imaginary part prevails.

The $\cal PT$ symmetry of the wave functions is broken for the eigenstates
with complex eigenvalues.
Figures~\ref{fig:4a}~(c) and (d) show as examples the wave functions 
obtained for $\gamma = 0.5$ for the corresponding pair of complex conjugate 
eigenvalues $\kappa$. It can be seen that the 
real and imaginary parts are no longer even or odd functions, and therefore 
$\cal  PT$ symmetry is lost. As a consequence, the moduli of the wave functions
also  are 
no longer even functions of $x$. Thus we find that beyond the branch
point not only the $\cal PT$ symmetry of the wave functions is broken but also
that of the nonlinear Hamiltonian! 

It must be noted, however, that
eigenstates with complex eigenvalues, in spite of  
solving the nonlinear 
eigenvalue problem (\ref{GPEdimless}), do not obey the 
corresponding {\em time-dependent} nonlinear Schr\"odinger equation.
For complex eigenvalues the modulus squared of the wave functions grows or decays proportional to $\exp(-2{\rm Im}(\kappa^2) t)$, and so does
the nonlinear term in  (\ref{GPEdimless}). Therefore strictly speaking multiplying
these eigenstates  with the usual time evolution factor
$\exp(i\kappa^2t)$ only captures the {\em onset} of the 
temporal evolution of the two modes, for times ${\rm Im}(\kappa^2) t \ll 1$.  
If one wished to obtain the
exact time evolution of these solutions in the region beyond the branch point 
one would have to solve the time-dependent nonlinear Schr\"odinger equation
(see subsection \ref{temporal_evolution}). 
In figure~\ref{fig:4a}(c) the mode with positive imaginary part
of $\kappa$ is the one which decays, as expected it is more 
strongly localized 
in the trap with loss, while the mode with negative imaginary part in
figure~\ref{fig:4a}(d) is the one
which grows and is more strongly localized in the trap with gain.

\begin{figure}[tb]
\begin{center}
\includegraphics[width=1\textwidth]{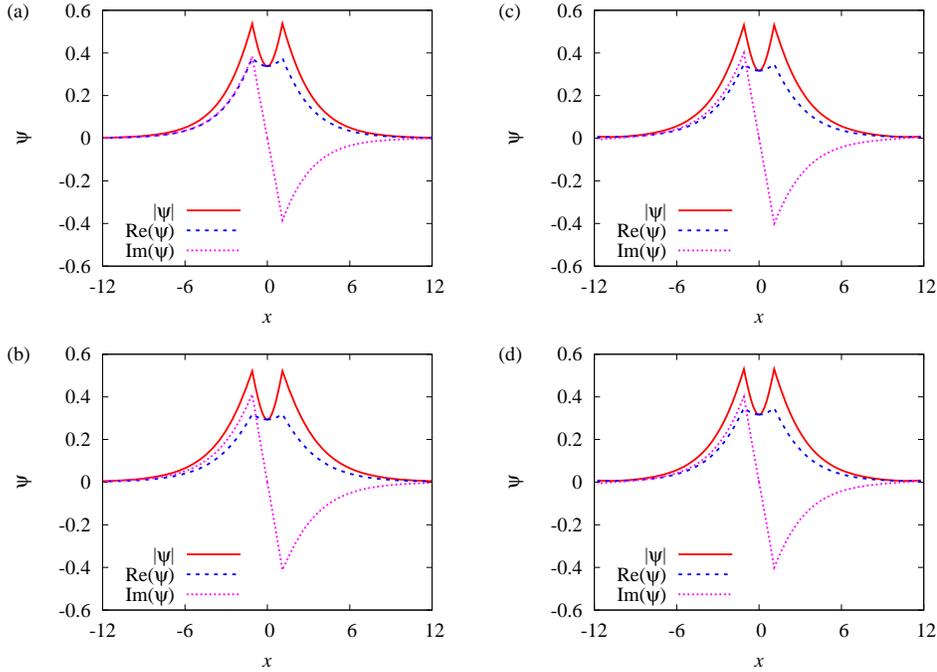}
\end{center}
\caption{\label{fig:8}
Wave functions for $a=2.2$ and $g = 0.5$ and values of $\gamma$ close to the exceptional point:
ground state (a) and excited state (b) for $\gamma = 0.395$, ground state (c)
and excited state (d) for $\gamma = 0.398$. While for $\gamma = 0.395$ 
still small differences can be recognized, in particular in the imaginary part, 
the wave functions practically conincide at $\gamma = 0.398$.
 }
\end{figure}

\begin{figure}[tb]
\begin{center}
\includegraphics[width=0.6\textwidth]{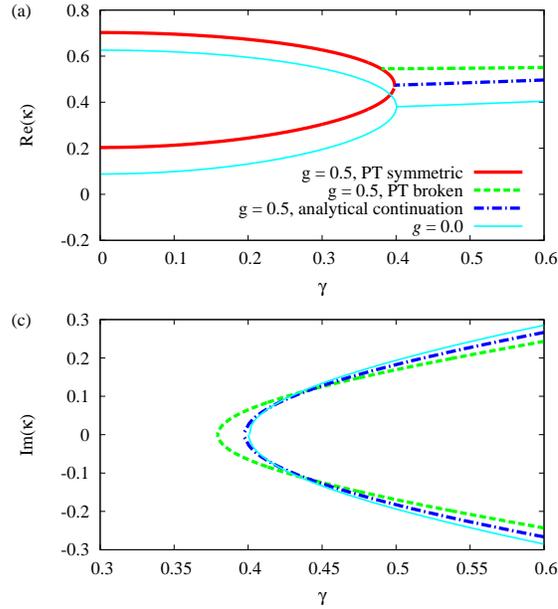}
\end{center}
\caption{\label{fig:10}
Eigenvalues of the nonlinear Hamiltonian with $a=2.2$ for
$g=0.5$ including the analytical continuation as
functions of the loss/gain parameter: (a) real part, (b) imaginary
part of $\kappa$ (unless zero). Now a pair of complex conjugate eigenvalues
also emerges at the exceptional point.  
 }
\end{figure}

\begin{figure}[ht]
\begin{center}
\includegraphics[width=1\textwidth]{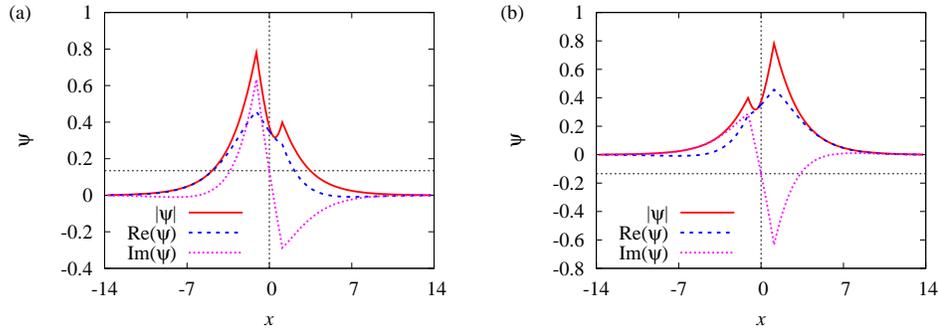}
\end{center}
\caption{\label{fig:11}
Wave functions of the eigenstates of the nonlinear Hamiltonian 
belonging to the complex eigenvalues of the analytical
continuation for $g=0.5$ and $a=2.2$ 
and a loss/gain parameter of  $\gamma = 0.5$. 
The eigenfunctions are not $\cal PT$ symmetric. The black dotted
lines have been drawn to illustrate that Im$(\Psi(0)) \ne 0$.
 }
\end{figure}

To demonstrate that at $\gamma_{\rm cr}$ the two eigenstates with real
eigenvalues indeed
coalesce, we have plotted their wave functions in figure~\ref{fig:8} 
for $a=2.2$, $g=0.5$, close to the critical value of $\gamma_{\rm cr}\approx 0.4$. While slightly below the critical value minor deviations of the 
wave functions are still visible, in particular in the imaginary parts,
no differences can be recognized anymore at $\gamma = 0.398$.
Thus  also for strong nonlinearity at $\gamma_{\rm cr}$  
we find the properties characteristic of an exceptional point, 
i.e. the coalescence of both eigenvalues and eigenfunctions.

\subsubsection{Analytical continuation of the nonlinear Hamiltonian}

The fact that at the branch point two real solutions coalesce 
without giving rise to two solutions with complex eigenvalues
contradicts the behaviour typical of exceptional points. 
Obviously these solutions cannot be found by solving the
nonlinear Gross-Pitaevskii equation in its form (\ref{GPEdimless}),
but require an analytical continuation of the nonlinear Hamiltonian
beyond the critical point $\gamma_{\rm cr}$. The reason is that the
nonlinear term $g|\Psi|^2$ is a nonanalytic function, and some
care has to be taken when analytically continuing the Hamilitonian
beyond the exceptional point.

In the $\cal PT$ symmetric regime up to $\gamma_{\rm cr}$ 
we have $\Psi^\ast(x) = \Psi(-x)$. Therefore
on the way to the bifurcation point we can write the nonlinearity
for the $\cal PT$ symmetric states in the form 
$g|\Psi(x)|^2 \equiv g \Psi(x) \Psi(-x)$. 
This function can be continued analytically. In the numerical calculation
the additional condition $\int \Psi(x) \Psi(-x) dx = 1$ must be introduced
to fix the phase of the nonlinearity in the $\cal PT$ broken
regime. In the (then) six-dimensional root search also ${\rm Im}(\Psi(0))$ must
be varied. As a result we find two more complex conjugate solutions 
that emerge from the coaelescing states, see figure~\ref{fig:10}.
The real and imaginary parts of the eigenfunctions and their moduli
are shown in figure~\ref{fig:11}. Obviously the states are not 
$\cal PT$ symmetric, and no longer possess vanishing imaginary part
at the origin. 

\subsubsection{Temporal evolution}\label{temporal_evolution}

\begin{figure}[tb]
\begin{center}
\vspace*{0.5cm}
\includegraphics[width=0.6\textwidth]{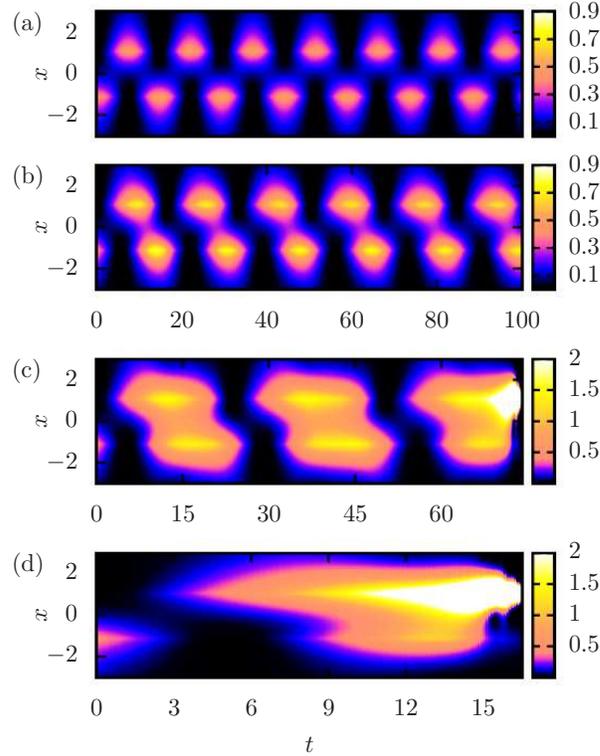}
\end{center}
\caption{\label{fig:9}
Temporal evolution of a superposition of the two stationary solutions
for $a=2.2$ and $g = 0.5$ and different strengths of the loss and gain
parameter:
(a) $\gamma = 0$, (b) $\gamma = 0.2$, (c) $\gamma = 0.275$, 
(d) $\gamma = 0.3$. 
 }
\end{figure}

We finally take  a look at the time evolution of a superposition of the
two stationary eigenstates  
in the $\cal PT$ symmetric regime. 
While the time dependence of each individual state is given
by $\exp(i\kappa^2 t)$, the time evolution of a superposition 
can no longer be obtained by simply multiplying the individual states
by their time evolution factors. The reason is again
that the nonlinear term in the Gross-Pitaevskii equation becomes
time-dependent, and therefore to determine the temporal evolution of 
an initial superposition the time-dependent Gross-Pitaevskii
equation has to be solved. We have done this numerically by applying
the split-operator technique, where in the short-time approximation
the time evolution operator can be split into products of exponentials
of the kinetic and potential terms of the (nonlinear, time-dependent) 
Hamiltonian,
and the time propagation is calculated by successively switching
between position and momentum space.  

In figure~\ref{fig:9} we show the results for the time evolution of
$|\Psi(x,t)|^2$ of the initial superposition of the ground and the
excited state $\Psi(x,0)=(\psi_1(x) + \psi_2(x))/\sqrt{2}$,  
for the nonlinearity $g=0.5$ and different values of the 
loss/gain parameter. It can be seen that, even with the nonlinearity
present, for small $\gamma$ we have the behaviour typical of $\cal PT$ symmetric systems,
i.e. the decrease of the beat length as the loss and gain parameter is growing
and the appearance of time intervals in which the probability
density in both wells is simultaneously large, or almost completely
extinguished. 

In figure~\ref{fig:9}~(d), for the value of $\gamma = 0.3$
the $\cal PT$ symmetry obviously is broken: even though we start
with a high probability density in the well with loss, as time
proceeds there is no beating, and the well with gain wins. Note that this value of $\gamma$ 
is smaller than the value of $\gamma \approx 0.38$ where the two complex eigenvalues
bifurcate from the ground state branch and smaller than 
the critical value $\gamma_{\rm cr} \approx 0.4$ where the two states coalesce. The reason is that
the norm of the state constantly varies with time, and with it the size of the
nonlinear term in the Gross-Pitaevskii equation. This effectively
corresponds to a variation of $g$ in the stationary Gross-Pitaevskii
equation, and acts as if one would constantly be moving through 
different eigenvalue diagrams. Therefore for a given value of $g$, 
the transition to the
$\cal PT$ broken phase of the condensate  can set in at values of 
$\gamma$ smaller than those of the bifurcation and the exceptional point
for that value of $g$ .

\section{\label{concl}Conclusions and outlook}

We have analyzed the simple quantum mechanical 
model of a Bose-Einstein condensate
in $\cal PT$ symmetric delta-function double traps by solving the 
stationary Gross-Pitaevskii equation first for negligible nonlinearity 
and then for finite nonlinearity. For vanishing nonlinearity we find that
the eigenvalues and eigenstates exhibit the same behaviour as in an
analogous optical system of two $\cal PT$ symmetric coupled wave guides
studied by Klaiman et al. \cite{klaiman08a}.
There exist two eigenvalues which merge at a critical value of the 
loss/gain parameter in a branch point, up to the branch point the 
corresponding two eigenstates are $\cal PT$ symmetric, they coalesce
at the branch point, beyond the branch point the eigenvalues become 
complex conjugate, and the $\cal PT$ symmetry of the wave functions
is broken. 

For finite nonlinearity we also find two branches of real eigenvalues
which merge in an exceptional point. We have the important result 
that the wave functions of the stationary states 
remain $\cal PT$ symmetric even if the nonlinearity is present. 
As a consequence their moduli are even functions,
and therefore the nonlinear state-dependent 
Hamiltonian selects as solutions exactly
such states which make itself $\cal PT$ symmetric. We also find a 
branch of two complex conjugate eigenvalues for which   
the $\cal PT$ symmetry of the wave functions is broken, and 
with it that of the nonlinear Hamiltonian. 

A surprising result is that, with the nonlinearity present, 
the branches of complex conjugate eigenvalues do not bifurcate 
from the exceptional point, but emerge at a smaller value of the loss/gain 
parameter from the eigenvalue branch of the ground state.
This implies that there exist ranges of the loss/gain parameter where
four eigenstates exist. 
This change in the branching scheme of a $\cal PT$ symmetric nonlinear
system is in agreement with the results found by Graefe et al.
\cite{graefe10} in their investigations of a Bose-Hubbard dimer.

The puzzle that at the exceptional point no branches of complex
eigenvalues are born could be solved by analytically continuing the 
nonlinear term in the Gross-Pitaevskii equation, and  we could 
show that indeed two more complex conjugate solutions are born at this point. 

The temporal evolution of an initial superposition of the two 
stationary states shows, even for nonvanishing nonlinearity,
the typical beating patterns 
known from $\cal PT$ symmetric systems, but we have found that
the breakdown of $\cal PT$ symmetry occurs at values of 
the loss/gain parameter smaller than those values where the two
branches of complex conjugate eigenvalues bifurcate from
the ground state or where the exceptional point appears. 
This is due to the time variation 
of the nonlinear term in the Gross-Pitaevskii equation.

We have considered the 
case of an attractive nonlinearity 
but found that the same behaviour occurs for repulsive nonlinearity.

Our investigations can be extended in several directions.
First it will be a worthwhile enterprise to investigate $\cal PT$ symmetric
Bose-Einstein condensates in realistic double well potentials
\cite{ananikian06}, in one or more dimensions, and 
possibly pin down physical parameters where the breaking of $\cal PT$ symmetry
could be observed in a real experiment.
Also, in addition to the nonlinearity resulting from the short-range contact 
interaction,
condensates with a long-range dipole-dipole interaction \cite{lahaye09b}
could be considered.  It would also be interesting  whether
for the nonlinear $\cal PT$ symmetric Hamiltonian considered in the present
paper similar to the work of Jones \cite{jones08}
the construction of a metric operator is possible with respect to
which the nonlinear Hamiltonian is quasi-Hermitian. 
Furthermore it would be desirable to find simple matrix models
which show the unusual branching scheme of the eigenvalues 
found for finite nonlinearity.
  
\ack
We thank Eva-Maria Graefe and Miloslav Znojil for helpful comments.\\

\appendix
\section{Gross-Pitaevskii equation}
Expressed in terms of physical quantities the Gross-Pitaevskii equation for a condensate in a $\cal PT$ symmetric
delta-functions double well reads
\begin{eqnarray}\label{GPE}
-\frac{\hbar^2}{2m}\frac{d^2}{dx^2}\Psi(x) &-&\big[{(V_0+i\Gamma)}\delta(x+b) + {(V_0-i\Gamma)}\delta(x-b)\nonumber\\
 &-& G|\Psi(x)|^2\big]\Psi(x) = \mu \Psi(x) \,.
\end{eqnarray}
The depths of the double wells is determined by $V_0$, the
size of the gain and loss terms by $\pm \Gamma$. The amplitude of the
nonlinear term which arises from the contact interaction between the condensate
atoms is given by the quantity $G$ and is proportional to the scattering 
length of the condensate atoms,
$\mu$ is the chemical potential. Measuring lengths in units of $L=\hbar^2/(2mV_0)$ and energies
in units of $E_0 = 2mV_0^2/\hbar^2$, brings (\ref{GPE})  into the dimensionless
form given in  (\ref{GPEdimless}).

\section*{References}
\bibliographystyle{unsrt}

\end{document}